\documentstyle[preprint,aps,psfig]{revtex}

\begin{document}

\tightenlines

\title{Linking Rates of Folding in Lattice Models of Proteins with
Underlying Thermodynamic Characteristics }
\author{D.K.Klimov and D.Thirumalai }

\address{Institute for Physical Science and Technology and 
Department of Chemistry and Biochemistry\\
University of Maryland, College Park, Maryland 20742}

\maketitle

\begin{abstract}
We investigate the sequence-dependent 
properties of proteins that determine the {\em
dual requirements of stability of the native state and its kinetic
accessibility } using simple cubic lattice models. Three
interaction schemes are used to describe the potentials between
nearest neighbor non-bonded beads. We
show that, under the simulation conditions when the native basin of
attraction (NBA) is the most stable, there is an excellent correlation
between folding times \(\tau _{F}\)  and the dimensionless parameter 
\(\sigma _{T} = (T_{\theta } -
T_{F})/T_{\theta }\), where \(T_{\theta }\) is the collapse
temperature and \(T_{F}\) is the folding transition temperature. 
There is also a significant
correlation between \(\tau _{F}\) and another dimensionless quantity 
\(Z=(E_{N}-E_{ms})/\delta \),
where \(E_{N}\) is the energy of the native state, \(E_{ms}\) is the
average energy of the ensemble of misfolded structures, and \(\delta
\) is the dispersion in the contact energies. 
An approximate relationship between \(\sigma _{T}\) and the
\(Z\)-score is derived, which explains the superior correlation seen
between \(\tau _{F}\) and \(\sigma _{T}\). {\em For two state folders} 
\(\tau _{F}\)  is linked to the {\em free energy
difference} (not simply  energy gap, however it is defined) 
between the unfolded states and the NBA.

\end{abstract}

\section{Introduction}

Natural proteins reach their native conformation in
biologically relevant time scale of about a second or less starting from an
ensemble of denatured conformations  \cite{Creightonbook}. 
The native state of proteins is also
stable (albeit marginally) under physiological conditions  
\cite{Creightonbook}. The underlying energy landscape of random
sequences is far too rugged \cite{Dill95,DillChan97,Bryn95} 
to be navigated in biologically relevant time scale. Thus, it is
believed that protein sequences have evolved so that the dual
requirements of stability and kinetic accessibility of their native states are
simultaneously satisfied. An important question that arises from this
observation is: What are the sequence dependent properties of proteins that
govern their foldability? The sequences that satisfy the above stated dual
requirements are considered to be foldable, and hence are biologically
competent. This and related questions have 
attracted considerable theoretical attention
over the last several years 
\cite{Dill95,DillChan97,Bryn95,Scheraga94b,Scheraga98,Skolnick96,Onuchic95,Cam93,Thirum95,KlimThirum96,KlimThirumPRL,Pande98}. 
Minimal protein models \cite{Dill95,DillChan97,Bryn95,KlimThirum96,Pande98}, 
which capture some but
not all the energetic balances in proteins, are particularly suited to
provide a detailed answer to the question posed here.

There have been three proposals in the literature, which have attempted to
identify the characteristics of sequences that give some insight into the
foldability. Below we briefly describe the three criteria following the
order of their appearance in the literature:

\begin{enumerate}

\item By using the random energy model (REM) 
as a caricature of proteins it
has been suggested \cite{Bryn95,Bryn89,Gold92} 
that foldable sequences have large values of \(T_{F}/
T_{g,eq}\) where \(T_{F}\) is the folding transition temperature and 
\(T_{g,eq}\) 
is an equilibrium glass transition temperature, which in the original REM
model is associated with the temperature at which the entropy vanishes. It
has been subsequently realized that in order to utilize this criterion in
lattice models \(T_{g,eq}\) has to be replaced by a kinetic glass transition
temperature \cite{Socci94}. 

\item Theoretical considerations and  
lattice and off-lattice model simulations show that for optimal 
foldable sequences the collapse transition temperature is
relatively close to \(T_{F}\) \cite{Cam93,Thirum95,KlimThirum96,Veit,Cam96}. 
In other words, sequences that
fold extremely rapidly have small values of 
\begin{equation}
\sigma _{T}=\frac{T_{ \theta } - T_{F}}{T_{\theta } } 
\label{sigmaT}
\end{equation}
where \(T_{\theta }\) is the temperature at which the polypeptide chain makes a
transition from the random coil state to a set of compact conformations. 
The characteristic temperatures \(T_{\theta }\) and \(T_{F}\) are
{\em equilibrium properties} 
that can be altered by not only changing the external
conditions, but also by mutations \cite{Thirum95,KlimThirum96}. 
The collapse transition temperature is, in
principle, measurable from the temperature dependence of the radius of
gyration, which can be measured using  small angle X-ray scattering 
(or neutron scattering) experiments.
We have shown that, depending
on the values of $\sigma _{T}$, the very nature of the folding kinetics can be
dramatically altered \cite{KlimThirum96,Veit}. 
In particular sequences for which \(\sigma _{T} \approx 0\) (here the
process of collapse and the acquisition of the native state is
{\em indistinguishable}) fold by two state kinetics. Such sequences are also
stable over a wider variety of external solvent conditions. On the
other hand, sequences, for which \(\sigma _{T}\) is relatively large,
exhibit more complicated kinetics \cite{Cam93,Thirum95}. 

\item
Finally it
has been argued that the "necessary and sufficient" conditions for a sequence
to be foldable is that there be a large energy gap (with dimensions
\(kcal/mol\)) or the native state be a
"pronounced" minimum in energy \cite{Sali94b}. The validity of this 
criterion, even for lattice models, has been questioned in several
articles  \cite{KlimThirum96,KlimThirumPRL,Cam96,Unger96,Grassberger97}.

\end{enumerate}

In general, sequences which fold
rapidly (small values of $\sigma _{T}$) are most easily generated by
performing some sort of optimization in sequence space. One popular way of
getting optimized sequences is to minimize the dimensionless \(Z\)-score 
\cite{Eisenberg,Gutin95} defined as 
\begin{equation}
Z = \frac{E_{N} - E_{ms}}{ \delta } 
\label{Zscore}
\end{equation}
where \(E_{N}\) is the energy of the native state, \(E_{ms}\) is the average
energy of the misfolded (or partially folded) states, and $\delta $ is the
dispersion in the contact energies. 
The purpose of this
paper is to investigate if the folding rates are correlated with the 
\(Z\)-score
in a manner similar to the correlation between folding times $\tau
_{F}$ and \(\sigma _{T}\) 
\cite{Cam93,KlimThirum96,KlimThirumPRL,Veit,Cam96}. We show, using a database 
of several lattice models of 
proteins, that there is a significant correlation between $\tau _{F}$ and the
{\em dimensionless quantity}  $Z$-score. The correlation, however, is not as
strong as that seen between $\tau _{F}$ and $\sigma _{T}$ at least in these
models. The rest of paper is organized as follows. In section II we present
the models and the computational protocol. In section III the stability of
the native state under the simulation conditions is established. The
correlations between $\tau _{F}$ and $\sigma _{T}$ 
and the $Z$-score are also discussed. We also establish an appropriate
relationship between \(\sigma _{T}\) and the \(Z\)-score. 
The paper is concluded in section IV with some additional
remarks.

\section{Methods}

\subsection{Lattice Models of Proteins }

We model a protein sequence as a
self-avoiding walk on a cubic lattice with the spacing \(a=1\) 
\cite{Dill95,Onuchic95}. 
A  conformation of a polypeptide chain  is 
given by vectors \(\{\vec{r}_{i}\}\), \(i=1,2...N\). 
The value of $N$ for three sequences is 36 and for the remaining
nineteen sequences, $N=27$. If, two
nonbonded beads \(i\) and \(j\) (\(|i-j| \ge 3\)) 
are nearest neighbors on a lattice, i.e., 
\(|\vec{r_{i}}-\vec{r_{j}}| = a \), they form a
contact. The energy of a conformation is given by the sum of
interaction energies \(B_{ij}\) associated with the contacts between
beads
\begin{equation} 
E = \sum _{i<j} \Delta (|\vec{r_{i}}-\vec{r_{j}}|-a) B_{ij},
\label{energy}
\end{equation} 
where \(\Delta \) is unity, when \(|\vec{r_{i}}-\vec{r_{j}}|=a\) and
is zero,
otherwise. We have used three forms for the contact matrix elements
\(B_{ij}\) which mimic the diversity of interactions between various
amino acids. 
Sixteen sequences with $N=27$ in this study have the contact matrix
elements \(B_{ij}\) obtained  from a Gaussian distribution \cite{KlimThirum96}
\begin{equation}
P(B_{ij})=\frac {1}{\sqrt{2\pi }B}\exp (-\frac{(B_{ij}-B_{0})^2}{2B^2})
\label{Bij}
\end{equation}
where \(B_{0}\) is the average attraction interaction and the
dispersion \(B\) gives the extent of diversity of the interactions among beads.
The energies for these sequences are  measured in terms of \(B\) which is
set to unity; \(B_{0}\) is taken to be  \(-0.1\). We will refer to
this interaction scheme as the random bond (RB) model. 
For three other sequences with $N=27$ and one sequence with $N=36$ 
\(B_{ij}\) are taken from Table III of 
ref. \cite{KGS}. We will refer to this interaction scheme
as the KGS model. 
A modified form of the Miyazawa-Jernigan potentials is
used for two $N=36$ sequences \cite{MJ85}. We will
denote this interaction scheme as the MJ model. 

Fifteen RB 27-mer sequences used  in this study are taken from our
previous work \cite{KlimThirum96}. 
An additional RB 27-mer sequence was included in our
database during the course of this work to expand the range of
\(\sigma _{T}\) values. Of the sixteen sequences nine
have maximally compact native states, while the remaining seven
sequences have non-compact native structures. 
Three KGS 27-mer sequences have identical maximally compact native
structures. Similarly three 36-mer sequences have identical
maximally compact native  conformations. 
The native conformation of 36-mer sequences is shown in
Fig. (1a).

For each sequence we perform Monte Carlo simulations (for details,
see ref. \cite{KlimThirum96}) 
and determined \(T_{\theta }\) and \(T_{F}\) using
multiple histogram technique \cite{Ferrenberg}, which is 
described in the context of protein folding  
elsewhere \cite{KlimThirum98,Brooks97}. Briefly, \(T_{\theta }\) is
associated with the peak of the specific heat \(C_{v}\) as a function
of temperature. Such estimates of  \(T_{\theta }\) 
coincide with the peak in the derivative of the temperature
dependence of the radius of gyration \(<R_{g}>\) \cite{KlimThirum98}. 
The folding
transition temperature is obtained from the peak of the fluctuations
in the overlap function, \(\Delta \chi \) \cite{Cam93}. These methods have been
successfully used to obtain the two characteristic equilibrium
temperatures for lattice, off-lattice, and all-atom models of
proteins \cite{Cam93,KlimThirum96,Veit,Okamoto,Brooks97}. 
In Fig. (1b) we show the temperature dependence of \(\Delta \chi
\), \(C_{v}\), and \(d<R_{g}>/dT>\) 
for the sequence whose native conformation is
shown in Fig. (1a).  
From the peaks of these plots we get \(T_{F} = 0.80\) and 
\(T_{\theta } = 1.14\), so that \(\sigma _{T} \) (see
Eq. (\ref{sigmaT})) for this sequence is 0.30. 

We also computed the \(Z\)-score (see Eq. (\ref{Zscore})) for the
twenty two sequences. The values of \(E_{ms} \) were calculated as 
\(E_{ms} = c<B>\), where $c$ is the number of contacts in the
misfolded structures and $<B>$ is the average contact energy for a
particular sequence. The dispersion \(\delta \) is determined from 
\(\delta ^{2} = <B^{2}> - <B>^{2}\), where \(<B^{2}>\) is the average
of the square of contact energies. In general, $c$ is equal to the number of
contacts in the native state which for maximally compact structures is
28 for $N=27$ and 40 for $N=36$. 
The kinetic simulations are done at sequence dependent temperatures 
\cite{KlimThirum96,Sali94b}, which were 
determined by the condition \(<\chi (T_{s})> = \alpha \). This
criterion for choosing \(T_{s}\) allows several sequences to be
compared on equal footing regardless of topology and the nature of
interaction potentials used. The value of
$\alpha =0.21$ is chosen so that \(T_{s} < T_{F}\) for all sequences. This
ensures that the native conformation or more precisely the native
basin of attraction  is the most dominant at \(T=T_{s}\). 

The folding times are calculated from the time dependence of the
fraction of unfolded molecules, \(P_{u}(t)\) \cite{Veit}. The
function  \(P_{u}(t)\) may be computed from the distribution of first
passage times. Operationally for every sequence an ensemble of
initial denatured conformations (obtained at \(T > T_{\theta }\)) is
generated. For each initial condition the temperature is reduced to
\(T_{s}\) and the dynamics is followed till the first passage time is
reached. Typically we generated between 200 to 500 independent 
trajectories in order to
reduce the  statistical error in determining \(\tau _{F}\) to about 5\%.

\section{Results}

\noindent 
\textbf{(a) Stability of the native basin of attraction: } Due to the
discrete nature of spatial and energetic representation the native state of
lattice models of proteins is a single microstate. In the coarse grained
energy representation every term in the energy function has Ising like
discreteness (see Eq. (\ref{energy})). Since these models 
represent a coarse grained caricature of
proteins it is useful to define a native basin of attraction (NBA) . This is 
necessary because the idea that the native state is a single microstate is
clearly unphysical. The native basin of attraction has a volume associated
with it. The larger such a volume is the smoother one expects the underlying
energy landscape to be. The probability of being in the NBA is defined
as \cite{Brooks97}
\begin{equation}
P_{NBA}(T)=\frac{\sum_{i} \delta (\chi _{i} \leq \chi _{NBA})
\exp^{-\frac{E_{i}}{kT}} }{\sum \exp^{-\frac{E_{i}}{kT}}}
\label{Pnba}
\end{equation}
where $\chi _{NBA}$ is the value of the overlap function at the folding
transition temperature \(T_{F}\), 
\(E_{i}\) is the energy of the conformation $i$, and 
\(\chi _{i}\) 
is the corresponding value of the overlap. The overlap function is
defined as \cite{Cam93,KlimThirum96}
\begin{equation}
\chi = 1 - \frac{1}{N^{2}-3N+2} \sum_{i\neq
j,j\pm 1} \delta(r_{ij} - r_{ij}^{N}),
\end{equation}
where \(r_{ij}\) is the distance between the  \(i\) and \(j\) beads
and  \(r_{ij}^{N}\) is the distance between the same beads in the
native conformation. 
According to Eq. (\ref{Pnba}) all
conformations with overlaps less than $\chi _{NBA}$ map onto the NBA,
which implies that a steepest descent quench would directly lead these
conformations  to the NBA.
The above definition of NBA is physically appealing. For all the
sequences \(T_{F}\), 
obtained from the peak of fluctuations in
the overlap function, nearly coincides with \(T_{F}\) 
determined using \(P_{NBA}(T_{F}) = 0.5\).

In order to demonstrate the stability of the NBA we have
calculated \(P_{NBA}(T_{s})\) for all the sequences. Recall that the sequence
dependent temperatures at which the simulations are performed are chosen so
that \(<\chi (T_{s}) > = 0.21\). In Fig. (2a) we show \(P_{NBA}(T_{s})\)
for the twenty two sequences. For all the
sequences the probability of being in the NBA exceeds 0.5, which 
implies that for
the simulation conditions employed here the {\em stability of the native
state is automatically ensured}. Among the nineteen 27-mer sequences
fifteen  are exactly the same ones 
as reported in our previous work \cite{KlimThirum96}, 
and the temperatures of the
simulations are identical to those used in our earlier studies. 
Thus, even in our earlier work the stability of
the NBA at the simulation temperatures has been guaranteed. 

\noindent 
\textbf{(b) Dependence of folding times on $\sigma _{T}$:} The folding times
for the 22 sequences considered here have been computed using methods
described in detail elsewhere \cite{Veit}. 
In Fig. (2b) we plot the dependence of 
$\tau _{F}$ on $\sigma _{T}$. This figure clearly shows that the folding
times correlate extremely well 
with $\sigma _{T}$ under conditions when the NBA is stable (see Fig. (2a)).
A relatively small change in $\sigma _{T}$ can lead to a dramatic increase in
folding times. For example, an increase in \(\sigma _{T}\) by a factor
of four results in  three orders of magnitude increase in \(\tau
_{F}\). This figure clearly shows that the dual requirements of
thermodynamic stability of the NBA and kinetic accessibility of the
NBA are satisfied for the 
sequences with relatively small values of $\sigma _{T}$. This verifies
the {\em foldability principle} which states that fast folding sequences with
stable native states have \(T_{F} \approx  T_{\theta }\) 
\cite{ThirumKlimWood}. 

There are two 
important points concerning the results presented in Fig. (2b): (1)
The excellent correlation shown in Fig. (2b) should be considered as
statistical i.e., we expect to find such dependence of $\tau _{F}$ on $\sigma
_{T}$ only if a number of sequences over a range of $\sigma _{T}$ is examined.
This implies that it is not possible to predict the relative rates of
folding for sequences whose $\sigma _{T}$ values are close. Such sequences are
expected to fold on similar time scales. (2) Foldable sequences with small
values of $\sigma _{T}$ reach the NBA over a wider range of external
conditions than those with moderate values of $\sigma _{T}$. Since not all
naturally occurring proteins fold rapidly it follows that there are proteins
with moderate values of $\sigma _{T}$ that reach their NBA by complex
kinetics \cite{Cam93,Thirum95}. 
This involves, in addition to the direct pathway to the native
state, off-pathway processes involving intermediates. Such sequences reach
the NBA by a kinetic partitioning mechanism \cite{Thirum95,ThirumKlimWood}.

\noindent 
\textbf{(c) Relationship between $\tau _{F}$ and $Z$-score: } The folding
times as a function of the $Z$-score for our database of sequences are shown
in Fig. (3a). It is clear that there is a significant correlation between the
two. However, 
the correlation here is not as good as that in Fig. (2b). The plausible
reasons are given in the next subsection in which the relationship between
$Z$-score and $\sigma _{T}$ is explored. It is tempting to think that because
the numerator of $Z$-score is some measure of the so called stability gap one
can conclude that folding rates are linked just to \(E_{N}-E_{ms}\). We show
below that this is not the case.

\noindent 
\textbf{(d) Relationship between $Z$-score and $\sigma _{T}$:} The significant
correlations between the folding times and $\sigma _{T}$ and $Z$-score suggest
that there might be a relationship between these two \textit{dimensionless
quantities}. We arrive at an approximate relationship between the two which
also explains the reasons for the superior correlation between $\tau _{F}$ and 
$\sigma _{T}$. 

The rationale for using $\sigma _{T}$ as a natural criterion that satisfies
the dual requirements of stability and kinetic accessibility is the
following \cite{ThirumKlimWood}. 
The transition from compact states to the native state at $T_{F}$
is usually first order, and neglecting the entropy associated with the native
state \(T_{F}\) is approximately given by \cite{Cam96}
\begin{equation}
T_{F} \approx \frac{|\delta E_{SG}|}{S_{NN}}
\label{estTF}
\end{equation}
where $\delta E_{SG}$ is roughly the stability gap and $S_{NN}$ is
the entropy of the (non-native) states  whose average energy is roughly 
$|\delta E_{SG}|$ above the NBA. If there is considerable 
entropy associated with the  NBA this has to
be subtracted from the denominator of Eq. (\ref{estTF}). 
Since our arguments do not really depend on
this, we ignore it here. The transition from the random coil states to the
collapsed states occurs at \(T_{\theta } \approx  D/k_{B}\), where $D$, is the
driving force, that places the hydrophobic residues in the core and the
polar residues in the exterior creating an interface between the compact
molecule and water. 

The entropy of the intervening non-native states is a function of $D$.
Consider the case of large $D$. In this case the polypeptide chain will
undergo a non-specific collapse into one of the exponentially large number of
compact conformations. This renders $S_{NN}$ extensive in $N$, where $N$ is the
number of amino acid residues in the polypeptide chain. This makes $T_{F}$
very low even for moderate sized proteins. In the opposite limit, when D is
small, there is not enough driving force to create a compact structure. Here
again $S_{NN}$ is extensive and as a result $T_{F}$ becomes low. Thus an
optimum value of $D$, which reflects a proper balance of local interactions
(leading to secondary structures) and long range interactions (causing
compaction and formation of tertiary structure) 
is necessary \cite{Go83} so that $S_{NN}$ be small
enough. This would make $T_{F}$ as large as possible without exceeding the
bound $T_{\theta }$. Thus, optimizing $\delta E _{SG}$, $S_{NN}$, and
$D$ (hence
$T_{\theta }$) leads to small values of $\sigma _{T}$, which therefore
emerges as a natural 
parameter that determines the folding rates and stability. 

Consider the spectrum of states of protein-like heteropolymers. It has been
suggested, using computational models \cite{Guo92} 
and theoretical arguments \cite{Gold92}, that
generically the spectrum of states consists of the NBA separated from the
non-native states by $\delta $E$_{SG}$. 
Above the manifold of non-native
states one has the ensemble of random-coil conformations. A lower limit of
the energy separating the random coil conformations and the non-native
compact structures is $\delta $, which is the dispersion in the energy of the
non-native states. (In lattice models \(\delta \) is associated with
the dispersion in the contact energies). 
Thus \(k_{B}T_{\theta } > \delta \). 
Assuming\footnote{Notice that the arguments do not depend on the
precise form of the density of states. All we require is that the
density of misfolded states has the functional dependence so that
\(S_{NN} = k_{B}ln\,\Omega _{NN}\) be positive. For example, if \(\Omega
_{NN}(E) \sim exp(E- \delta E_{SG})\), then \(S_{NN} \sim
k_{B}ln[sh(\delta/2\delta_{0})]\), where $\delta _{0}$ is a 
suitable constant. }
that the density of non-native structures in the energy range \(-\delta
/2 \le E-\delta E_{SG} \le \delta /2\) is \(\Omega _{NN} \simeq 
(E - \delta E_{SG})^{\alpha }\) 
(where \(\alpha \)
is an even integer) we get \(S_{NN} \approx k_{B} ln(\delta
/\delta _{0})\), 
where \(\delta _{0}\) is a sequence dependent constant. From these arguments it
follows that 
\begin{equation}
\frac{T_{F}}{T_{\theta } } < \frac {|Z|}{\frac{S_{NN}}{k_B}} 
\approx \frac{|\delta E_{SG}|}{\delta } \frac {1}{C \ln \frac {\delta }{\delta _{0}}} 
\end{equation}
if $\delta E_{SG}$ is identified with \(E_{N} - E_{ms}\). Thus, maximizing the
ratio \(T_{F}/T_{\theta }\) (or minimizing $\sigma _{T}$) is 
\textit{ approximately } equivalent 
to minimizing the ratio \(Z/(S_{NN}/k_{B})\). It is
perhaps the neglect of the entropy of the non-native states that leads to
the poorer correlation between \(\tau _{F}\) and $Z$-score as compared to
correlation between \(\tau _{F}\) and \(\sigma _{T}\) (see Figs. (2b)
and (3a)).

\noindent 
\textbf{ (e) Linking \(\tau _{F}\) to various definitions of the energy gap:}
Since the numerator of the $Z$-score is an estimate of \(\delta
E_{SG}\),  it might be tempting to conclude that there
is  relationship between \(\tau _{F}\) and the
associated energy gap. In the context of minimal models of proteins a number
of definitions of the "energy gap" have been proposed. It 
is useful to document these definitions:

\begin{enumerate}

\item \textit{Standard Energy Gap:} The time honored definition of the energy
gap  for any system (not consisting  of fermions)  
is \(\Delta _{P} = E_{N} - E_{1}\), where \(E_{N}\) is the
energy of the ground (native) state and \(E_{1}\) is that of the first excited
state. This definition is usually deemed inappropriate for protein-like
lattice 
models because a flip of one of the beads can lead to a trivial structural
change especially for non-compact native states. 
Such structures would belong to the NBA (see Eq. (\ref{Pnba})). 
In real models in water, 
even a casual flip of a residue could involve substantial solvent
rearrangements, resulting in significant energy (or enthalpy)  penalty. 

\item \textit{Compact Energy Gap: } 
The compact energy gap is \(\Delta _{CS} = E_{N}^{CS}-E_{1}^{CS}\) 
\cite{Sali94b},  
where \(E_{N}^{CS}\) and \(E_{1}^{CS}\) are the energies of the native
state and the first excited state respectively. The superscript $CS$ indicates
that these conformations are restricted to the ensemble of maximally compact
conformations. It has been shown that the ground states of many sequences
are non-compact \cite{KlimThirum96}. 
Furthermore, in several cases the lowest energy of the
maximally compact conformation is much greater than those of the manifold of
non-compact non-native structures \cite{KlimThirum96}. 
The correlation between \(\tau
_{F}\)  and \(\Delta _{CS}\) is, at best, poor (see Fig. (22) 
of ref. \cite{KlimThirum96}).

\item \textit{\ Stability gap: }The notion that $\delta E_{SG}$ should play an
important role in determining both the stability and the folding rates is
based on sound physical arguments \cite{Bryn95}. We believe that its close
relation to \(T_{F}\) makes the stability gap a very useful physical
concept. 

\item \textit{$Z$-score gap:} This is defined as 
\(\Delta _{Z} = E_{N}-E_{ms}\), which is the numerator of Eq. (\ref{Zscore}).
This is closely related to \(\delta E_{SG}\), and for practical purposes may be
identical. The precise value for \(E_{ms}\) depends on the given sequence, and
even 
the practitioner. The great utility of the $Z$-score is that one can use it as
a technical device to assess the efficiency of threading algorithm or for
generating sequences that are good folders \cite{Eisenberg,Gutin95}. 
For these two purposes the
precise values of \(E _{ms}\) 
do not appear to be very important. Since \(E_{ms}\)
can be altered freely the definition of $\Delta _{Z}$ is somewhat ambiguous.
In this article we have used the same definition of $E_{ms}$ for all
sequences, and this allows
us to assess the efficiency of the $Z$-score in determining folding kinetics.

We have tested the relationship between $\tau _{F}$ and \(\Delta _{Z}\) 
using the database of 22 sequences. A plot of the 
folding time for the 22 sequences as a function of \(\Delta _{Z}\) 
is given in Fig. (3b). We see a {\em very poor correlation 
between } \(\tau _{F}\)
{\em and} \(\Delta _{Z}\) for all sequences included in our database. 
There appears to be a
link between \(\tau _{F}\) and \(\Delta _Z\) 
for the three 27-mer sequences with the KGS
potentials but none for the three 36-mer sequences. 
The number of sequences with the KGS
interaction scheme and the modified MJ scheme is too small for meaningful
trend to be established. However, it is clear that 
the overall correlation between \(\tau _{F}\) and 
\(\Delta _{Z}\) is poor.

\end{enumerate}

\noindent 
\textbf{ (e) Probing the correlation between \(\tau_{F}\) and  
the free energy of stability: } The various energy gaps described 
above do not adequately correlate with \(\tau _{F}\). 
A plausible reason could be that the energy
gaps ignore the entropy of the chain in the denatured states. 
Here, we explore the idea that the
free energy of stability of the native state itself could be an indicator of
foldability. Consider a large number of two state
folders. In this case 
only the NBA and the ensemble of unfolded states which
have very little overlap with the native state are significantly
populated. From a physical point of
view, the appropriate equilibrium quantity that could correlate with
the folding rates is the free energy difference between the two
states. We consider the free energy of stability defined for two state
folders (with small values of \(\sigma _{T}\)) as
\begin{equation}
\Delta F _{U-N}= -k_{B}T_{s} ln K(T_{s})
\label{DeltaF}
\end{equation}
with the equilibrium constant  
\begin{equation}
K(T_{s}) = \frac{P_{NBA}(T_{s})}{1-P_{NBA}(T_{s})}
\end{equation}
where \(T_{s}\) 
is the simulation temperature and \(P_{NBA}(T)\) is given in 
Eq. (\ref{Pnba}). In experiments \(\Delta F_{U-N}\) 
should be replaced by \(\Delta G_{H_{2}O}\)
which gives the stability of the native state in the limit of zero denaturant
concentration.

In order to examine the dependence of \(\tau _{F}\) on 
\(\Delta F_{U-N}\) we
singled out the two state folders from our database. For these sequences
we computed \(\Delta F_{U-N}\) using Eq. (\ref{DeltaF}). 
In Fig. (4) we plot $\tau _F$ as
a function of \(\Delta F_{U-N}\). We do find important correlation
which approaches the quality of that 
shown in Fig. (2b). For the two state folders it is clear that 
\(\Delta F_{U-N}\) 
is a good estimate of \(k_{B}T_{F}\) and hence \(T_{\theta }\) (since 
\(\sigma _{T}\) is small). Thus the correlation seen in Fig. (4) is not
entirely unexpected.

\section{Conclusions}

The variations in 
folding times for a variety of sequences under conditions when the
native basin of attraction is the most populated can be understood in terms of
\(\sigma _{T}\) ( see Eq. (\ref{sigmaT}))
\cite{Thirum95,KlimThirum96}. Thus, the simultaneous
requirements of thermodynamic stability and kinetic accessibility are
satisfied for sequences for which \(\sigma _{T}\) is small. Such
sequences are foldable over a broad range of external conditions. 

It might be tempting to conclude that computation of \(T_{\theta }\)
and \(T_{F}\) for lattice models requires exhaustive simulations. This
is not the case. In order to get reasonably accurate estimate of 
\(T_{\theta }\) and  \(T_{F}\) by multiple histogram method we find
that, for most sequences, between 8 to 10 trajectories each with about
50 millions of Monte Carlo steps are sufficient. By comparison,  
reliable determination of folding kinetics time scales requires a few
hundred trajectories  at various temperatures.    
Thus, \(\sigma _{T}\)
is a useful criterion for designing fast folding sequences. By
contrast, notice that when a $Z$-score optimized sequence is
generated, its thermodynamics (as well as kinetics) 
is {\em a priori} unknown. A separate set of
simulations has to be performed at various temperatures in order to 
obtain  its thermodynamics.

There is a significant correlation
between \(Z\)-score and the rates of folding. This correlation is
not nearly as good as the one between \(\tau _{F}\) and \(\sigma
_{T}\). The  connection between \(\tau _{F}\) and the 
\(Z\)-score suggests that this could arise because the entropy of the
non-native states (or more precisely the entropy difference between
the non-native states and the native basin of attraction) is not taken
into account in the \(Z\)-score. More importantly, the \(Z\)-score does
not appear to be easily measurable making its experimental validation
difficult, if not impossible. 

There appears to be no useful predictive relationship between the
various energy gaps and the folding times. In seeking a
correlation involving energetics and entropy of the unfolded and
folded states we have found that for two state folders  {\em the free energy
of stability of the native state with respect to unfolded states}
correlates well with the folding times. Note that 
this quantity includes the entropies of
the NBA and the unfolded states. The correlation between the
folding time and \(\sigma _{T}\), and with the free energy of stability
for two state folders can be verified experimentally.

\acknowledgments

This work was supported in part by  a grant from 
the National Science
Foundation (through grant number  CHE96-29845).

\newpage

\begin{figure}

\noindent 
{\bf Fig. (1)} (a) 
The conformation of the native state of a 36-mer MJ sequence. This
sequence is  {\bf SQKWLERGATRIADGDLPVNGTYFSCKIMENVHPLA}, 
where we have used the one letter representation of the amino acids. 
This conformation is the lowest
energy conformation in the native basin of attraction. (b) Temperature
dependence of the fluctuations in the overlap function $\Delta \chi
$ (solid line), specific heat \(C_{v}\) (dotted line), and
the derivative of the radius of gyration with respect to temperature 
\(d<R_{g}>/dT\) (dashed line) for the sequence whose native state
is displayed in Fig. (1a). The scale for \(C_{v}\) and \(d<R_{g}>/dT\) 
is given on the right.  
The collapse temperature \(T_{\theta }\), obtained  from the larger peak of
specific heat \(C_{v}\) curve, is found to be 1.14.
It is seen that \(T_{\theta }\) is very close to the temperature at which
\(d<R_{g}>/dT\) reaches maximum (at 1.19). The two peaks in
\(d<R_{g}>/dT\), with the smaller one coinciding with the location of
the maximum in \(\Delta \chi\), suggest that from a thermodynamic
viewpoint a three state description is more appropriate for this sequence.
The value of \(T_{F}\), 
which is associated with the peak of \(\Delta \chi \), is 0.80.
Therefore, for this sequence collapse and folding transition
temperatures are separated by a large interval,  
and \(\sigma _{T}\) (=0.30) is consequently large.  
For this sequence 
the value of \(T_{F}\) obtained 
from the condition \(P_{NBA}(T_{F}) = 0.5\) is 0.79, 
which nearly coincides with the peak position of \(\Delta \chi \). In
majority of the sequences we only observe one peak in \(C_{v}\) and
\(d<R_{g}>/dT\) . Hence, it is necessary to introduce independent
order parameters to determine \(T_{F}\).

\noindent 
{\bf Fig. (2)} (a) 
The values of the probability of being in the native basin of
attraction \(P_{NBA}\) 
at the sequence dependent simulation temperatures \(T_{s}\)
for the database
of 22 sequences considered in this study. The horizontal dotted line
corresponds to \(P_{NBA} = 0.5\). This figures shows that at the simulation
temperatures \(P_{NBA}\) exceeds 0.5 which implies that the stability
criterion is automatically satisfied. (b) Plot of the folding times 
$\tau _{F}$ as a function of $\sigma _{T}$ for the 22 sequences. 
This figures shows that
under the external conditions when the NBA is the most populated there is a
remarkable correlation between $\tau _{F}$ and $\sigma _{T}$. {\em 
The correlation
coefficient is 0.94.} It is clear that over a four orders of magnitude
of folding times \(\tau _{F} \approx
\exp (-\sigma _{T}/\sigma _{0})\) where $\sigma _{0}$ is a
constant. In both panels the
filled and open circles are for the RB and KGS 27-mer models, 
respectively. The open squares are for $N = 36$.

\noindent 
{\bf Fig. (3)} (a) The dependence of \(\tau _{F}\) on the $Z$-score. 
There is a
significant correlation between the folding times and the $Z$-score. Since the
scales for the $Z$-score depend on both the interaction scheme and the length
of the sequence it is hard to fit the data for all 22 sequences. If we
restrict ourselves to the 16 sequences in the RB model, we find that the
correlation coefficient is 0.70, which is not nearly as good as in Fig. (2b).
(b) Plot of $\tau _{F}$ as function of $\Delta _{Z}$ 
which is the energy gap that
appears in the numerator of Eq. (\ref{Zscore}). 
This figure clearly shows that there is
no correlation between $\tau _{F}$ and the $\Delta _{Z}$. 
The various symbols are the same as in Fig. (2b).

\noindent 
{\bf Fig. (4)} 
This figure shows, for the two state folders only, the dependence
of $\tau _{F}$ on the free energy of stability $\Delta F_{U-N}$ of
the NBA with respect to denatured states. 
Notice that $\Delta F_{U-N}$ is not an energy
gap. It includes the entropies of the folded and unfolded states explicitly
and is obtained from the equilibrium constant between the unfolded states
and the NBA at the simulation temperature \(T_{s}\).

\end{figure}

\newpage 

\begin{center}
\begin{minipage}{10cm} 
\[
\psfig{figure=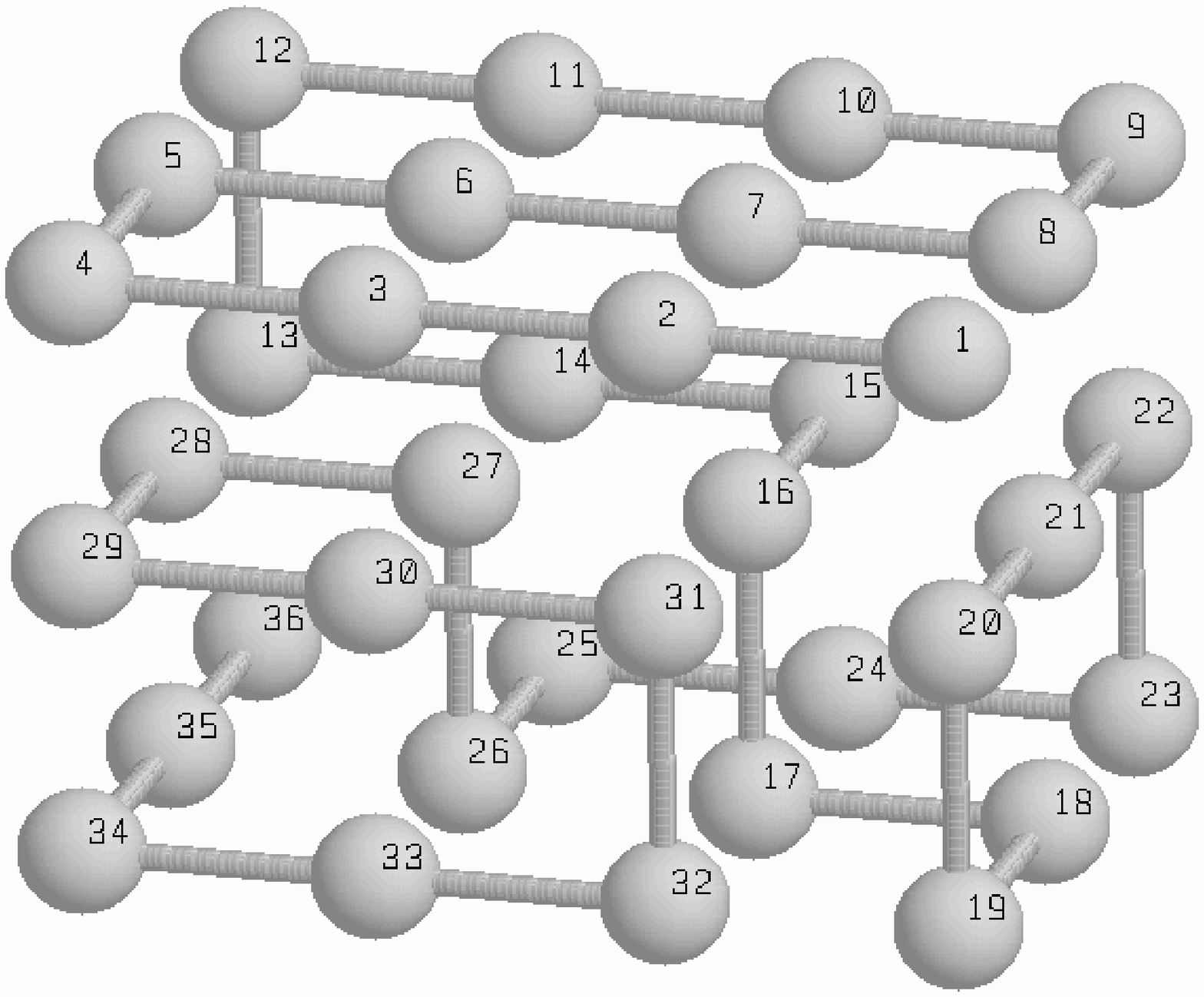,height=10cm,width=10cm}
\]
\end{minipage}

{\bf Fig. 1a} 
\end{center}

\begin{center}
\begin{minipage}{15cm}
\[
\psfig{figure=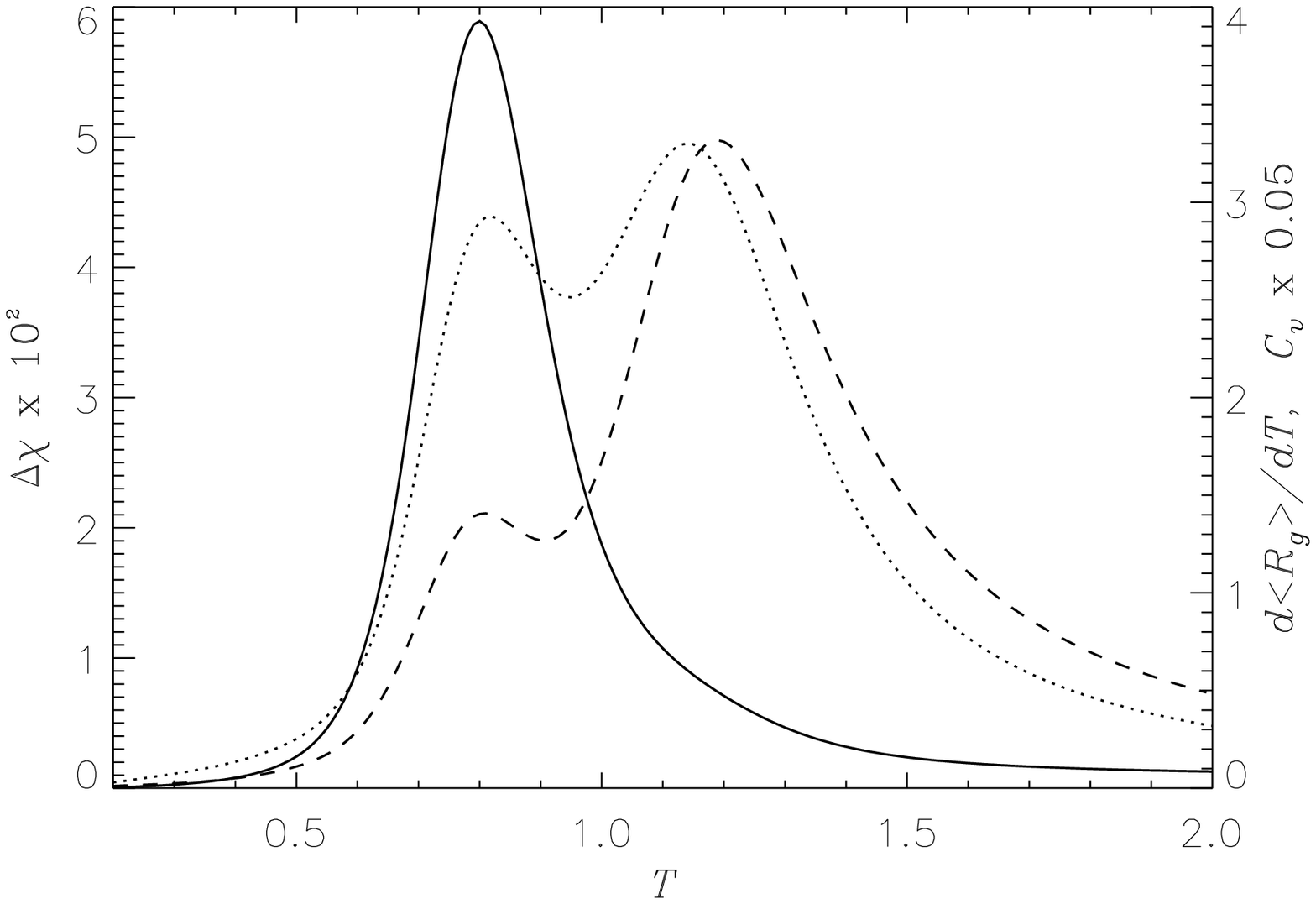,height=9cm,width=12cm}
\]
\end{minipage}

{\bf Fig. 1b} 
\end{center}

\newpage

\begin{center}
\begin{minipage}{15cm}
\[
\psfig{figure=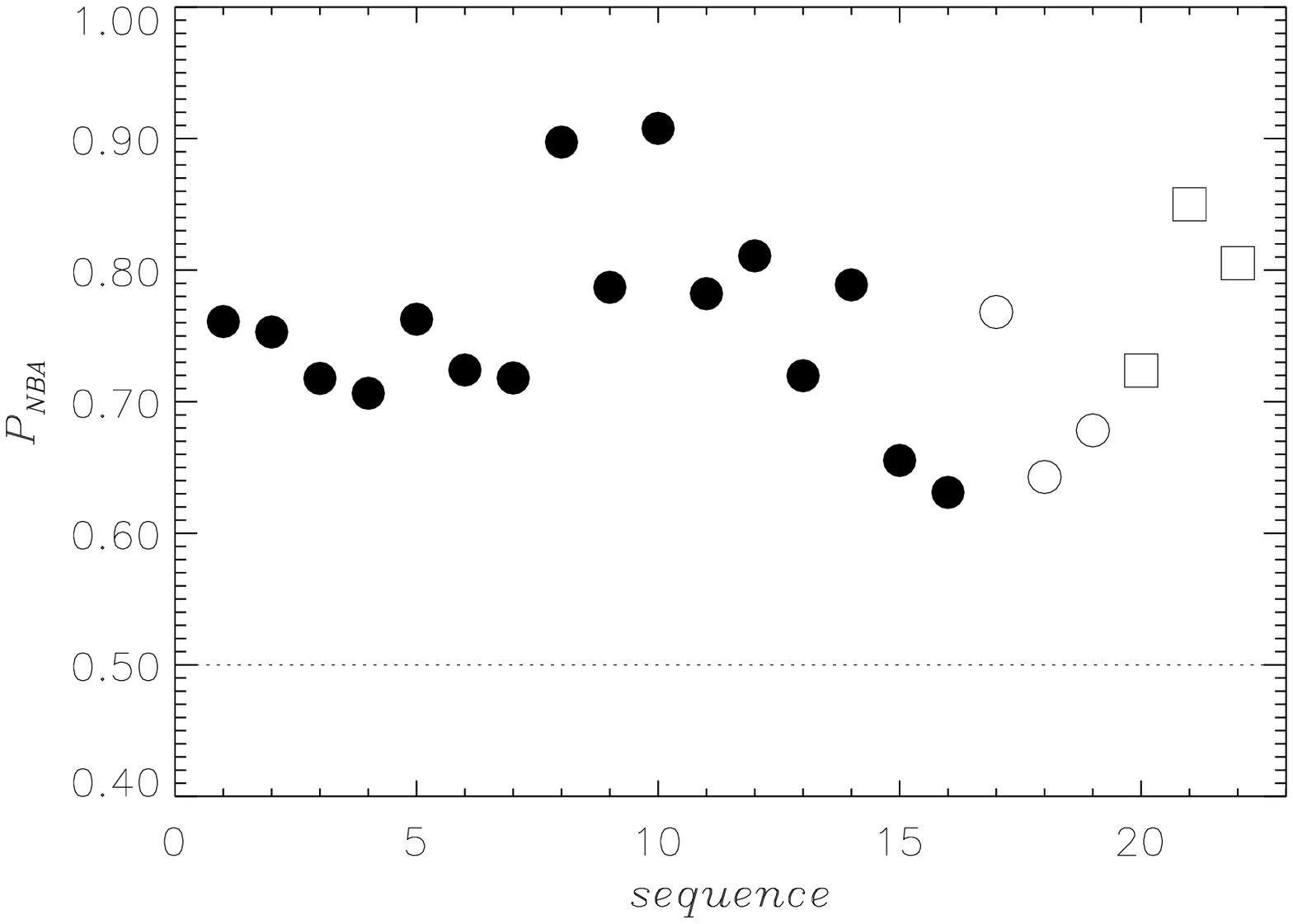,height=9cm,width=12cm}
\]
\end{minipage}

{\bf Fig. 2a} 
\end{center}

\begin{center}
\begin{minipage}{15cm}
\[
\psfig{figure=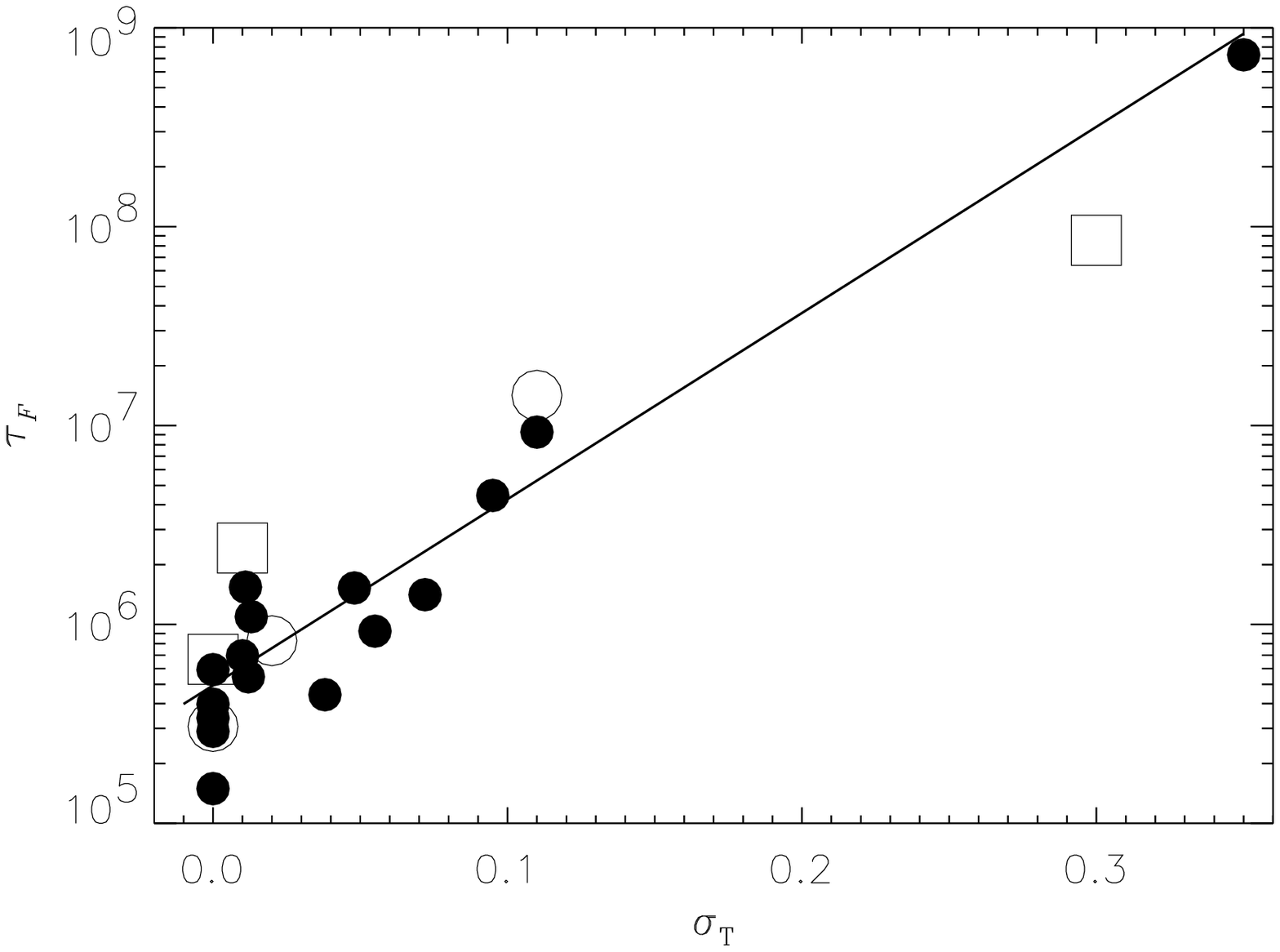,height=9cm,width=12cm}
\]
\end{minipage}

{\bf Fig. 2b} 
\end{center}

\newpage

\begin{center}
\begin{minipage}{15cm}
\[
\psfig{figure=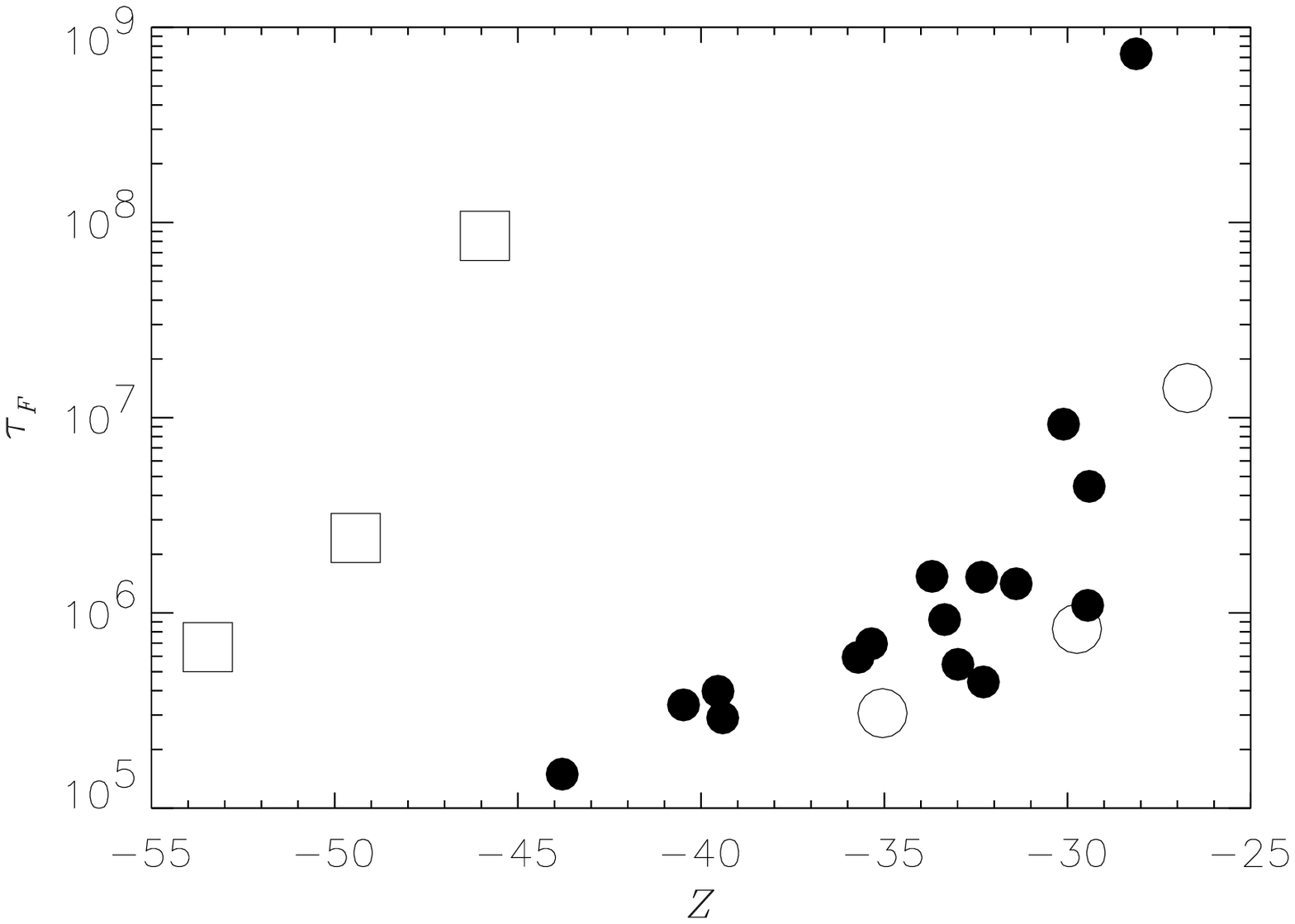,height=9cm,width=12cm}
\]
\end{minipage}

{\bf Fig. 3a} 
\end{center}

\begin{center}
\begin{minipage}{15cm}
\[
\psfig{figure=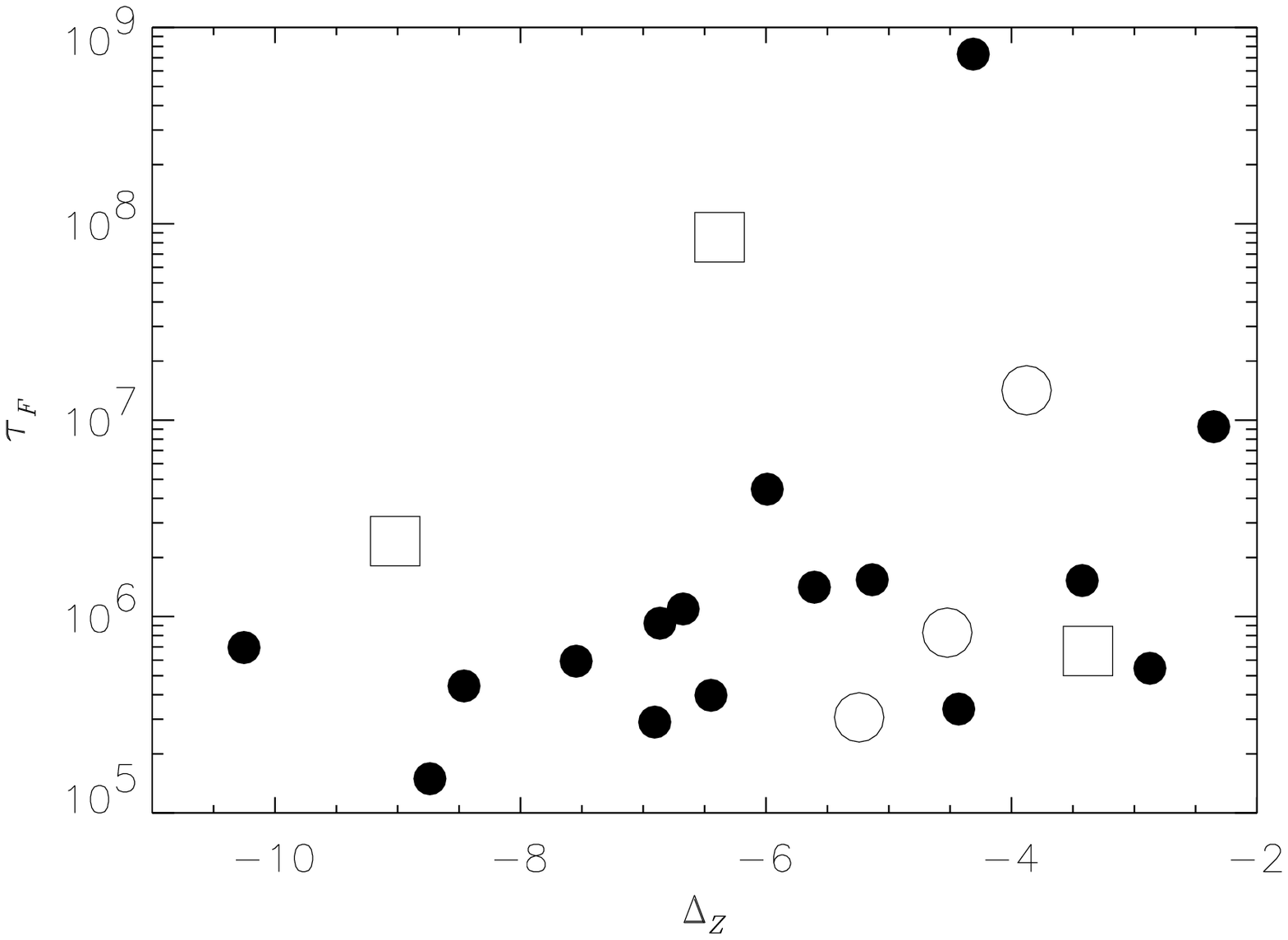,height=9cm,width=12cm}
\]
\end{minipage}

{\bf Fig. 3b} 
\end{center}

\newpage

\begin{center}
\begin{minipage}{15cm}
\[
\psfig{figure=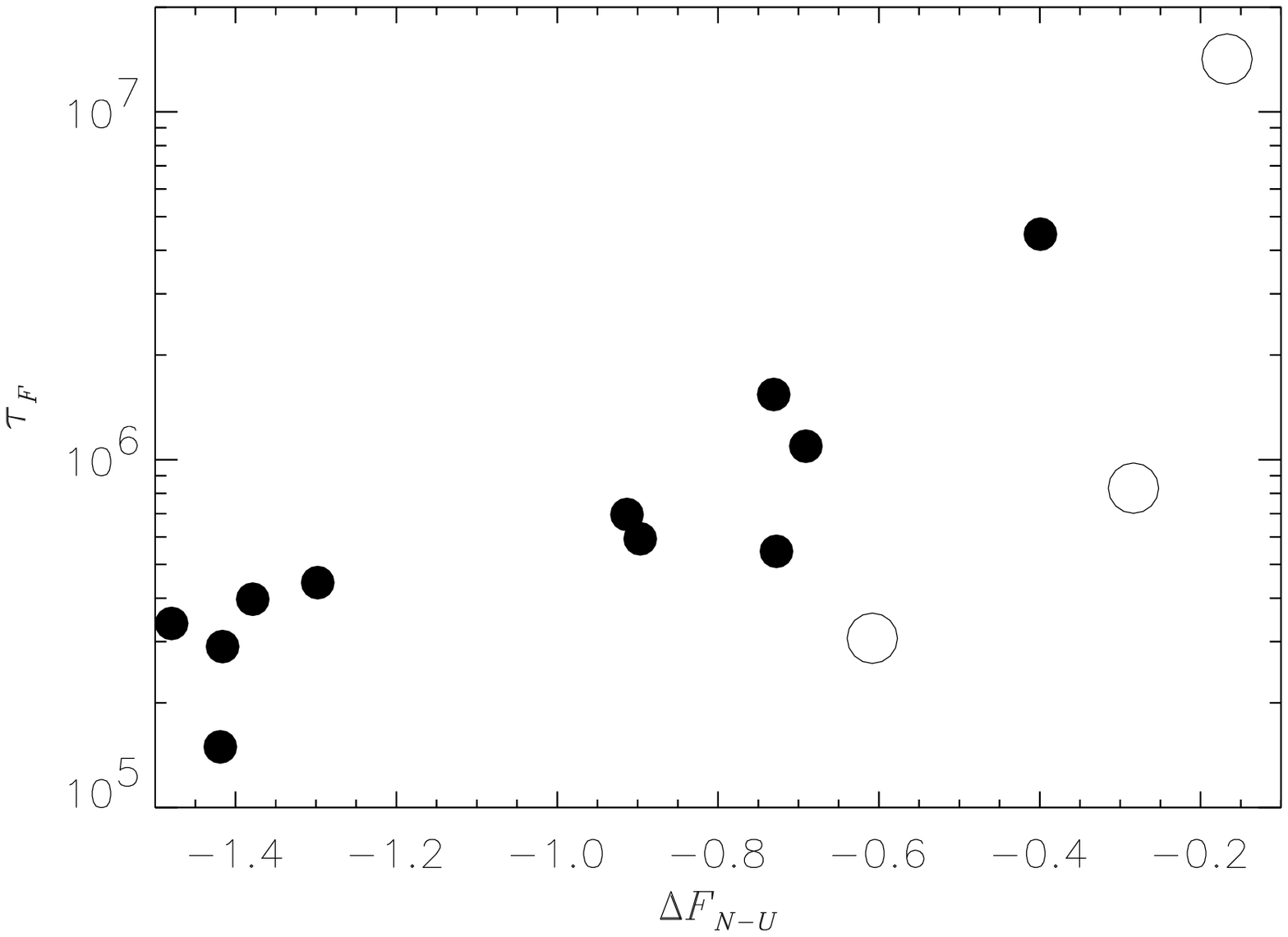,height=9cm,width=12cm}
\]
\end{minipage}

{\bf Fig. 4} 
\end{center}


\begin{references}


\bibitem{Creightonbook} T. E. Creighton, {\em 
Proteins: Structures and
Molecular Properties} (W.H. Freeman \& Co., New York, 1993).  


\bibitem{Dill95} K.A. Dill, S. Bromberg, K. Yue, K.M.  Fiebig, D.P. Yee, 
P.D. Thomas and  H.S. Chan, Protein Sci.  {\bf 4}, 561 (1995). 

\bibitem{DillChan97} K.A. Dill and  H.S. Chan, Natur. Struct. Biol. {\bf 4},
10 (1997). 

\bibitem{Bryn95} J.D. Bryngelson, J.N. Onuchic, N.D. Socci, and P.G. Wolynes,
Proteins Struct. Funct. Genet. {\bf  21}, 167 (1995). 
 
\bibitem{Scheraga94b} M.-H. Hao  and H.A. Scheraga, J.  Phys. Chem. 
{\bf 98}, 9882 (1994). 

\bibitem{Scheraga98} S. He  and H.A. Scheraga, J.  Chem. Phys.  
{\bf 108}, 287 (1998). 

\bibitem{Skolnick96} A. Kolinski, W.  Galazka, and  J. Skolnick,  
Proteins Struct. Funct. Genet.  {\bf 26}, 271 (1996). 


\bibitem{Onuchic95} N.D. Socci and J.N. Onuchic, J. Chem. Phys. {\bf
103}, 4732 (1995). 

\bibitem{Cam93}  C.J. Camacho and   D. Thirumalai, 
Proc. Natl. Acad. Sci. USA {\bf 90}, 6369 (1993).

\bibitem{Thirum95} D. Thirumalai, J. de Physique I {\bf 5}, 
1457 (1995). 


\bibitem{KlimThirum96} D.K. Klimov and  D. Thirumalai,  
Proteins Struct. Funct. Genet. {\bf 26}, 411 (1996). 

\bibitem{KlimThirumPRL} D.K. Klimov and  D. Thirumalai, Phys. Rev. Lett. 
{\bf 76}, 4070 (1996). 

\bibitem{Pande98} 
V.S. Pande, A.Yu. Grosberg, T. Tanaka, and D.S. Rokhsar, 
Curr. Opin. Struct. Biol. {\bf 8}, 68 (1998). 


\bibitem{Bryn89} J.D. Bryngelson and P.G. Wolynes, J. Phys. Chem. {\bf
93}, 6902 (1989). 

\bibitem{Gold92} R.A. Goldstein, Z.A. Luthey-Schulten and  P.G Wolynes,  
Proc. Natl. Acad. Sci. USA {\bf 89}, 4918 (1992).

\bibitem{Socci94} N.D. Socci and J.N. Onuchic,  J. Chem. Phys. {\bf
101}, 1519 (1994). 

\bibitem{Veit} T. Veitshans,  D.K. Klimov, and D. Thirumalai,  
Folding \& Design  {\bf 2}, 1 (1996). 

\bibitem{Cam96} C.J. Camacho and  D. Thirumalai, Europys. 
Lett. {\bf 35}, 627 (1996). 

\bibitem{Sali94b} A. Sali, E. Shakhnovich, and M. Karplus, 
J. Mol. Biol. {\bf 235}, 1614 (1994).




\bibitem{Unger96} R. Unger and J. Moult, J. Mol. Biol. {\bf 259}, 988
(1996). 



\bibitem{Grassberger97} U. Bastolla, H. Frauenkron, E. Gerstner,
P. Grassberger, and W. Nadler, cond-mat/9710030. 

\bibitem{Eisenberg} J.U. Bowie, R. Luthy, and D. Eisenberg,
Science  {\bf 253}, 164 (1991). 

\bibitem{Gutin95} A.M. Gutin, V.I. Abkevich, and E.I. Shakhnovich,
Proc. Natl. Acad. Sci. USA {\bf 92}, 1282 (1995). 



\bibitem{KGS} A. Kolinski, A. Godzik, and J. Skolnick,  
J. Chem. Phys. {\bf 98}, 7420 (1993). 

\bibitem{MJ85}  S. Miyazawa and R.L. Jernigan, Macromolecules 
{\bf 18}, 534 (1985). 

\bibitem{Ferrenberg} A.M. Ferrenberg and  R.H. Swendsen, 
Phys. Rev. Lett. {\bf 63}, 1195 (1989). 

\bibitem{KlimThirum98} D.K. Klimov and D. Thirumalai,
Folding \& Design {\bf 3}, 127 (1998). 

\bibitem{Okamoto} U.H.E. Hansmann, M. Masuya,  and Y. Okamoto, 
Proc. Natl. Acad. Sci. USA  {\bf 94}, 10652 (1997). 


\bibitem{Brooks97} Z. Guo and   C.L. Brooks III,  
Biopolymers {\bf 42},  745 (1997). 

\bibitem{ThirumKlimWood} D. Thirumalai, D.K  Klimov, and  S.A. Woodson,
Theor. Chem. Acct. {\bf 1}, 23 (1997). 

\bibitem{Go83} N. Go, Ann. Rev. Biophys. Bioeng.  {\bf 12}, 183 (1983). 

\bibitem{Guo92} Z. Guo, D. Thirumalai, and J.D.  Honeycutt,
J. Chem. Phys. {\bf 97}, 525 (1992). 







\end{references}
\end{document}